\documentclass[usenatbib]{mn2e} % 2-column version
\usepackage{natbib,float,graphicx,color,longtable}
\usepackage{amsmath,amsfonts,amssymb}
\usepackage{subfigure}
\usepackage{epsf}
\newcommand{\aap}{A\&A}
\newcommand{\aj}{AJ}
\newcommand{\apj}{ApJ}
\newcommand{\apjl}{ApJ}
\newcommand{\pasp}{PASP}
\newcommand{\pasj}{PASJ}
\newcommand{\mnras}{MNRAS}
\newcommand{\etal}{\text{et al. }}%{\textsl{et al. }}

\def\ion#1#2{{\rm #1}~{\rm #2}}
\def\urltilda{\kern -.15em\lower .7ex\hbox{\~{}}\kern .04em}
\def\urldot{\kern -.10em.\kern -.10em}
\def\urlhttp{http\kern -.10em\lower -.1ex\hbox{:}\kern -.12em\lower
0ex\hbox{/}\kern -.18em\lower 0ex\hbox{/}}

\begin{document}
\bibliographystyle{mn2e}

\title{A New Insight into the Classification of Type Ia Supernovae}

\author[V. ~Arsenijevic]{Vladan Arsenijevic\thanks{E-mail: arsenije@sim.ul.pt}\\
SIM, Faculdade de Ci\^encias da Universidade de Lisboa, Campo
Grande, C8, 1749-016, Lisbon, Portugal}
\maketitle

% due to elusive nature

\begin{abstract}
Type Ia Supernovae (SNe~Ia) spectra are compared regarding the
coefficient of the largest wavelet scale in their decomposition. Two distinct
subgroups were identified and their occurrence is discussed in light of use of
SNe~Ia as cosmological probes. Apart from the group of normal SNe,  
another trend characterised by intrinsically redder colours is consisted of 
many different SN events that exhibit diverse properties, including the
interaction with the circumstellar material, the existence of specific
shell-structure in or surrounding the SN ejecta or super-Chandrasekhar mass
progenitors. 
Compared with the normal objects, these SNe may violate the
standard width-luminosity correction, which could influence the cosmological
results if they were all calibrated equally, since their fraction among SNe Ia
is not negligible when performing precision cosmology. 
Using largest wavelet scale coefficient in combination with long-baseline
$B-I$ colours, we show how to disentangle SN intrinsic colour from the part
that corresponds to the reddening due to dust extinction in the host galaxy in
the SALT2 colour parameter $c$, discussing how the intrinsic colour
differences may explain the different reddening laws for two subsamples.
There are wavelength intervals for which the measured largest scale
coefficient is invariant to the additional extinction applied to a spectrum.
Combination of wavelet coefficients measured in different wavelength intervals
can be used to develop a technique that allows for estimation of extinction.

\end{abstract}

\begin{keywords}
supernovae: general  --- cosmology: observations
\end{keywords}

\bigskip\bigskip

\section{Introduction}

Type Ia Supernovae, widely accepted to be thermonuclear explosions of
carbon/oxygen white dwarfs, represent a homogeneous class of
stellar objects, both spectroscopically and photometrically. The possibility to
standardise the peak luminosity of SNe~Ia, no matter how far they are, makes
them efficient cosmological tool. Unfortunately, small differences among
SNe~Ia directly translate onto errors on the cosmological parameters. We have
reached a point where these errors are no more small for our goals. In order to
obtain a better accuracy of distance determination, we aim to
model the particular differences in a way to reduce as much as possible both
systematic and statistical errors on the cosmological parameters.
Indeed, several different techniques for fitting SN light-curves have been
widely used (e.g. \citealt{jha07,guy07,conley08}). Besides, some spectral
indicators, such as line ratios (\citealt{nugent95}) or pseudo-equivalent widths
($EW$) (\citealt{hachinger06,bronder08}), are found to correlate with absolute
magnitudes, thus providing a basis for an independent calibration method of the
SN luminosity. A detailed study on the selection of global spectral
indicators can be found in \citealt{bailey09}. However, the spectroscopic
diversity of SNe~Ia is multidimensional (e.g.
\citealt{benetti05,wang09,branch09}), and a whole range of reported subclasses
suggests that our understanding of models that correspond to these events is
still not complete. Particularly, the observed spectral differences between
SNe~Ia can be linked to variations in their effective temperatures, which are
associated with different amount of $^{56}$Ni produced in the explosions, i.e.
different luminosities (see e.g. \citealt{mazzali07}).

In many cases the high-resolution SN spectroscopy is rare and, therefore, we
cannot track the possible signature of the interaction between SN ejecta and
surrounding circumstellar material (CSM), normally exhibited by emission lines
in the spectra. It is demonstrated in this paper that the coefficient that
corresponds to the largest scale in the wavelet decomposition of SN spectrum
(for details see \citealt{arsenijevic08}), coupled with the SALT2 colour
(\citealt{guy07}), is a parameter that provides additional information on SN
subclassification, distinguishes SNe that show spectral peculiarities and most
likely indicates different SN progenitor scenario and/or explosion mechanism.
The wavelet inverse of the above-mentioned coefficient depicts the
overall shape of the SN spectrum. Depending on the wavelength interval, the
measured coefficient can be more sensitive to reddening. A method based on
comparison of two largest scale wavelet coefficients from different wavelength
intervals that indicates peculiar SNe, also those with anomalous extinction, is
described. 

Both intrinsic SN colour and host galaxy dust extinction are entangled
into single SALT2 colour parameter $c$ (\citealt{guy07}). The unfolding of
these effects which provides a possibility to estimate the SN host galaxy
reddening, $E(B-V)_\textrm{host}$, will be also discussed. 

\section{Data set}

The discrete wavelet transform is applied to a sample of 73 nearby SNe,
mostly found in the literature and presented in Table~\ref{nearbysnelist}. In
addition, an unusually bright SN 2003fg (also known as SNLS-03D3bb,
\citealt{howell06}), observed at intermediate redshift $z = 0.244$,
is also considered. All the low-resolution spectra ($\sim$600)
have been deredshifted, but no 
reddening correction has been made. For most of the analysis our attention is 
restricted to the rest-frame wavelength interval
$3400<\lambda<7000\;\textrm{\AA}$ and epoch within $[-10,+10]$ days relative
to $B$-maximum. We perform then the Discrete Time Wavelet Transform ($DTWT$)
of the spectra following the prescription given in \citealt{arsenijevic08}.
The decomposition is produced using Daubechies' extremal phase wavelets with 4
vanishing moments (D8).
All SNe in our sample have relatively good photometry, and in order to apply a
consistent procedure, the SALT2 light-curve fitter is used for all supernovae to
obtain the photometric parameters. However, the spectra employed in this study 
come from various sources and are not quite homogeneous, thus there are some
limitations and uncertainties on spectra imposed by reduction issues that may
affect the SN colour terms and certainly are not subject to intrinsic SN
properties.

\section{Method and Results}
\label{secres}

\begin{figure*}
\begin{minipage}[b]{0.5\textwidth}
\centering 
\subfigure[]{ \label{wltcoeff15color}
\includegraphics[width=8cm,angle=-90]{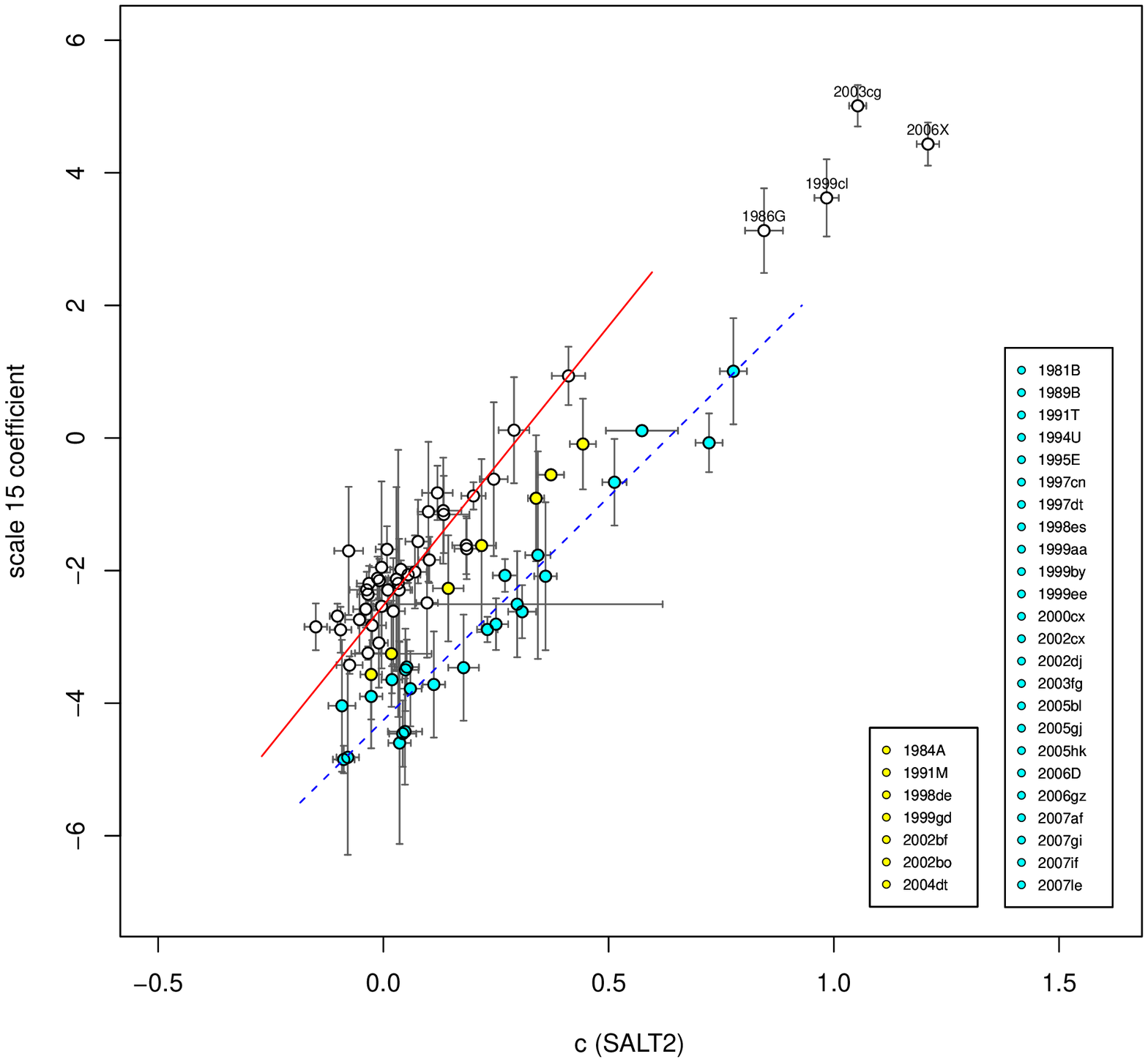}}
\end{minipage}% 
\begin{minipage}[b]{0.5\textwidth} 
\centering 
\subfigure[]{ \label{wltcoeff15colorb}
\includegraphics[width=8cm,angle=-90]{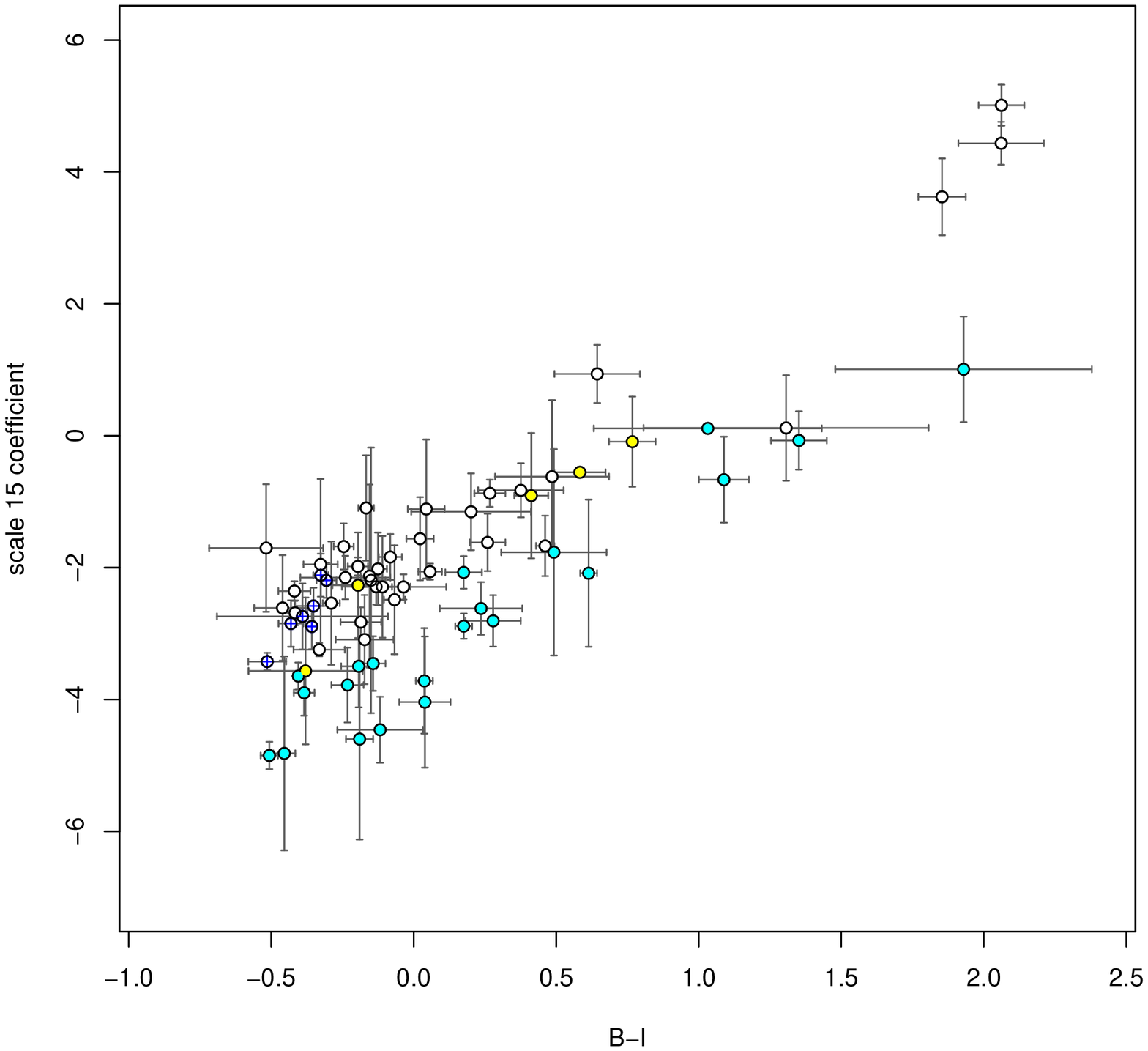}}
\end{minipage}
\caption[]{\small\sl (a) Mean value of the
wavelet coefficient from the largest scale for different
spectra of nearby SNe~Ia. The solid line corresponds to the best fit to
``normal'' SN data, while the dashed line describes the other SN class
(see text for details). (b) Same as in (a), but using a long-baseline
$B-I$ colour instead of the SALT2 $c$ parameter. Seven normal SNe that suffer 
negligible host extinction are identified as crossed symbols.} 
\label{wltcoeff15colors} 
\end{figure*}

SN~Ia spectra are decomposed into different wavelet scales following the
procedure described in \citealt{arsenijevic08}. In order to normalise
better the SN spectra, the authors do not include the contribution from scale
15, i.e. the coarsest wavelet scale in the spectral decomposition. Indeed, after
measuring the featureless inverse of the wavelet coefficient from scale 15 at
central wavelengths of $B$ and $V$ bands on our spectra, we see that a
contribution from the largest scale to the $B-V$ colour can be expressed as
$-0.0021\times \textrm{wltcoeff}_\textrm{15}$, which gives a maximum amount of 
$\sim0.01$ mag when applied to a range spanned by the coefficients,
between $-5$ and $0$ (see Fig.~\ref{wltcoeff15color}). The inverse of the
wavelet from the largest scale may be regarded as a black-body
curve or a sort of continuum. On the other hand, we deal here with the
spectral decomposition that does not map exactly to the underlying physics. In
other words, there is no true physical meaning of each component (scale) in a
wavelet decomposition; 
however one is able to measure the weight of each scale in terms of energy, as
the wavelet power spectrum suggests. Further, a spectral analysis that is based
on the wavelet decomposition certainly assures consistency of the method.
Nevertheless, plotting the wavelet coefficient from the largest scale as a
function of the SALT2 colour parameter, i.e. long baseline $B-I$ colour
in Fig.~\ref{wltcoeff15colorb}, leads to an interesting correlation.

The robust regression fit in Fig.~\ref{wltcoeff15color} reads
$c\textrm{(SALT2)}=
(0.406\pm0.025)+(0.125\pm0.009)\times
\textrm{wltcoeff}_\textrm{15}$, with the standard deviation of fit
$\hat{\sigma}=0.25$, computed as the median absolute deviation of the residuals
divided by the constant $0.6745$. 
% that makes the estimate unbiased for the normal distribution.
In order to reduce large data scatter, two perceivably distinctive SN trends are
identified, represented by solid and dashed lines in
Fig.~\ref{wltcoeff15color}:
\begin{equation}
c_\textrm{(SALT2)}=
(0.300\pm0.025)+(0.112\pm0.012)\times \textrm{wltcoeff}_\textrm{15}
\label{sc15colorfit}
\end{equation}
and
\begin{equation}
c_\textrm{(SALT2)}=
(0.632\pm0.026)+(0.149\pm0.008)\times \textrm{wltcoeff}_\textrm{15}
\label{sc15colorfitodds}
\end{equation}
respectively, with the robust measures of spread equal to 0.09, i.e. 0.11.  
Similarly, in Fig.~\ref{wltcoeff15colorb}, the scale 15 coefficient is well
correlated with a long-baseline $B-I$ colour at $B$-band maximum
corrected for Galactic reddening assuming $E(B-I)= 2.35 E(B-V)$. To calculate
the uncertainty of the $(B-I)$ values, the photometric errors with the
uncertainties due to the fitting procedure are combined. 
%({mccall04}?). 

A trend described by solid line in Fig.~\ref{wltcoeff15color} is referred
to as ``normal'' SNe. On the other hand, the earlier mentioned SN 2003fg is
identified as a member of SN subclass (represented by dashed line),
that includes objects whose colour appears to be redder than the typical
SN~Ia, according to the fits from Eqns.~\ref{sc15colorfit} and
\ref{sc15colorfitodds}. Looking closely to the members of this SN group,
represented by filled circles, there is some doubt whether these differences
reflect the different explosion scenario or distinct progenitor channels,
whether it can be attributed to the existence of (overdense) shell-structure in
or around the SN ejecta, the interaction with the circumstellar material,
massive progenitors, or finally, to different viewing angles of asymmetric
envelopes, as will be discussed in subsequent section.

\begin{figure*}
\begin{minipage}[b]{0.5\textwidth}
\centering 
\subfigure[]{ \label{bicolor}
\includegraphics[width=8cm,angle=-90]{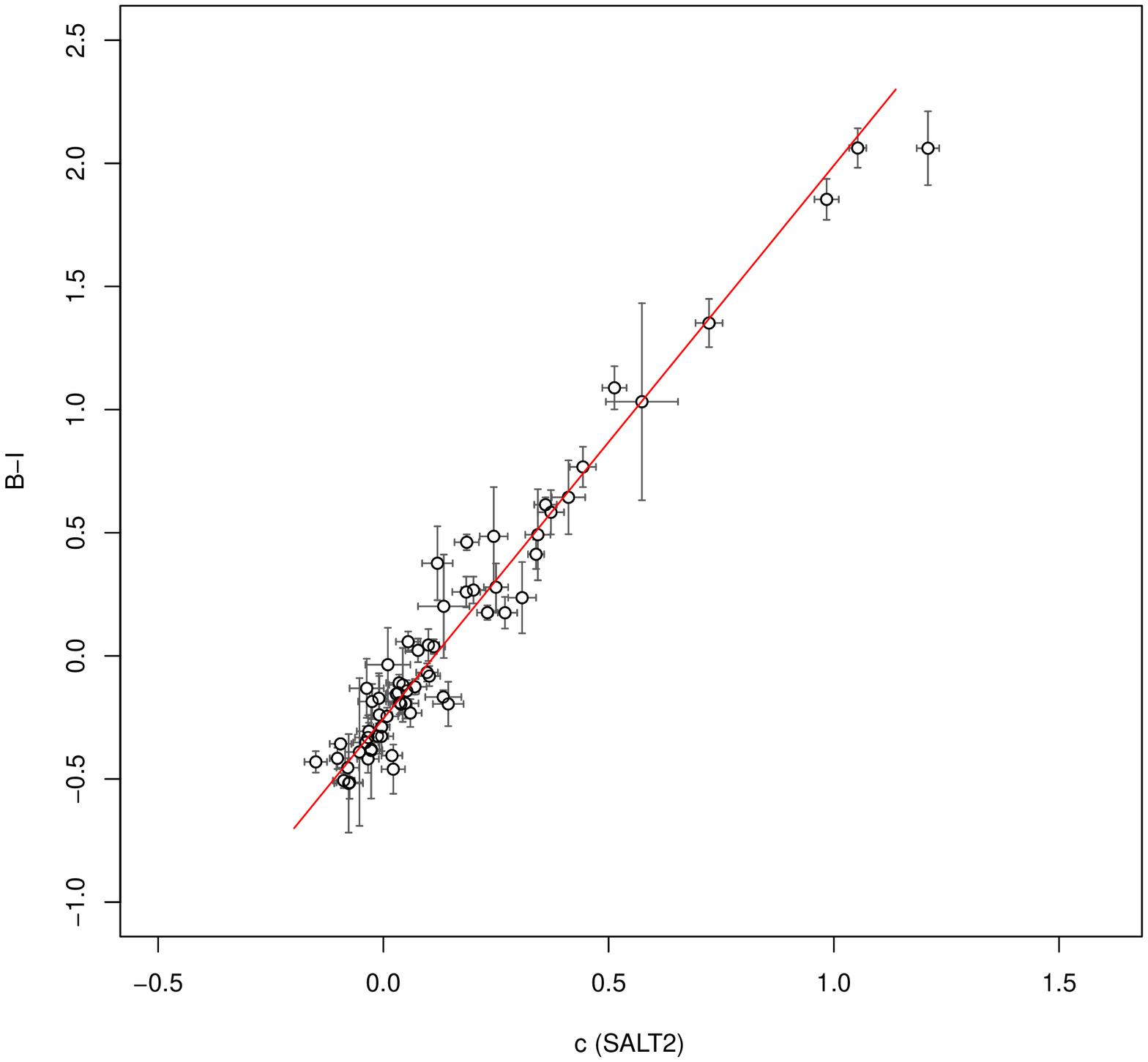}}
\end{minipage}% 
\begin{minipage}[b]{0.5\textwidth} 
\centering 
\subfigure[]{ \label{ebvhostcolor}
\includegraphics[width=8cm,angle=-90]{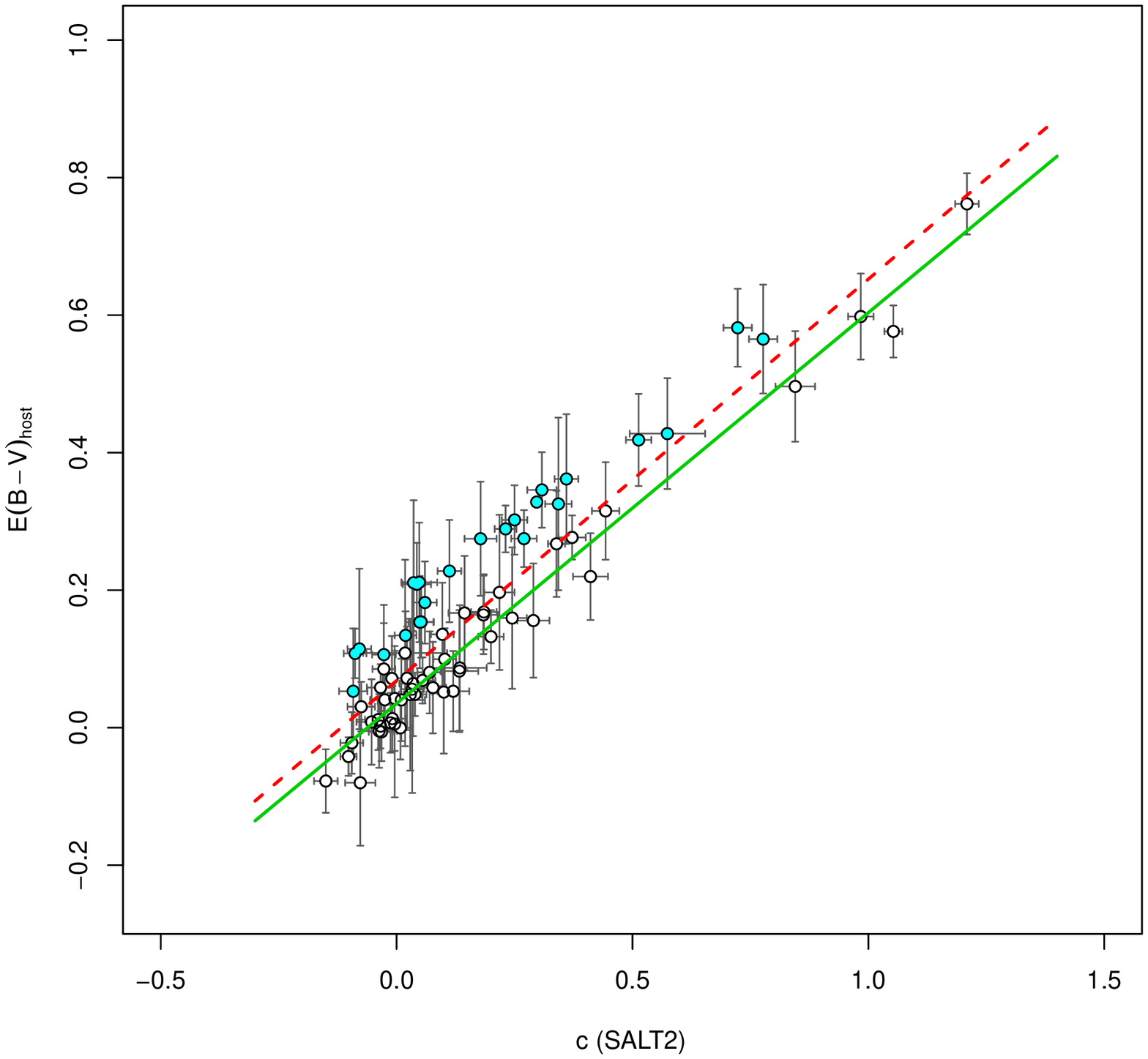}}
\end{minipage}
\caption[]{\small\sl (a) $B-I$ vs the SALT2 $c$ parameter. (b) Estimated host
galaxy reddening as a function of the SALT2 colour. Again, the filled circles
correspond to those in Fig.~\ref{wltcoeff15colors}. The best fit lines
for all/normal SNe are shown as dashed/solid lines.} 
\label{biebvcolors} 
\end{figure*}

Indeed, \citealt{blondin06} found that spectrum of SN 2003fg matches best with
the spectrum of SN 1989B at $t=+3.5$ days (with constraints on epoch or
redshift), an object also identified to belong to the same group.
\citealt{howell06} and \citealt{hillebrandt07} debate between the possible
explosion scenarios for SN 2003fg: a super-Chandrasekhar-mass progenitor
explosion or the asymmetric explosion of a Chandrasekhar-mass ($M_\textrm{Ch}$)
white dwarf? 
Something similar can be said for overluminous SN 2007if
(\citealt{scalzo10}) and for SN 2006gz (\citealt{hicken07}), both with
progenitor mass larger than $M_\textrm{Ch}$ and both surrounded by an envelope
of unburned carbon/oxygen.
Interestingly, SN 2009dc (\citealt{yamanaka09}), another SN with a
massive progenitor, seems\footnote{A value close to 1.5 for the wavelet
coefficient from the largest scale is measured from the available spectra from
\citealt{tanaka09}.} to be closer to normal SNe~Ia
in Fig.~\ref{wltcoeff15color}. 
%Unfortunately, there is no available spectral
%data below $4500\;\textrm{\AA}$ for this SN (at least to our knowledge),
%thus an additional check for classification is needed. 
Recently \citealt{silverman11} published 9 spectra of this SN that cover the
epochs from -7 days to +281 days and extend to $3500\;\textrm{\AA}$. An
additional check for classification of this object on the spectrum at -7 days
shows that the scale 15 coefficient for SN 2009dc is roughly $-4\pm 1$,
therefore this SN belongs to the second group, as was expected. 
SNe~Ia found to deviate from the main SN trend in Fig.~\ref{wltcoeff15colors},
whose colours are described by Eqn.~\ref{sc15colorfitodds}, may show
significant discrepancy
when a single parametrisation is applied, like the Phillips 
relation (\citealt{phillips93}) that relates a light-curve width to peak
luminosity, due to intrinsically different explosions. Also, as
noted in \citealt{wang05}, the circumstellar dust or a surrounding
gaseous envelope may substantially affect the extinction properties compared
to the interstellar dust. In other words, it is not wise to calibrate these SNe
as standard candles; their reddenings will be more likely overestimated, which
further implies that these objects are overcorrected to a higher luminosity,
that could bias the cosmological results. 
Moreover, a certain number of SNe fall within the rough range $-0.1<c<0.1$.
Interestingly, imposing this cut-off criterion in Fig.~\ref{wltcoeff15color},
the correlation between scale 15 coefficient and colour appears to be weak and 
the existence of distinct SN groups is no longer maintained.  
Nevertheless, one could select a subset of ``more standardisable'' SNe~Ia based
on the value of the largest wavelet scale coefficient.

\subsection{Intrinsic SN colour and host galaxy reddening}

As previously seen in Fig.~\ref{wltcoeff15colors},
the scale 15 coefficient is correlated with both the SALT2 colour and
long-baseline $B-I$ colour in a quite similar manner. Moreover, a strong
relationship between these latter two is shown in Fig.~\ref{bicolor}. The best
fit line corresponds to:
\begin{equation}
c_\textrm{(SALT2)}=(0.114\pm0.007)+(0.445\pm0.012)\times(B-I).
\label{cbi}
\end{equation}
In Fig.~\ref{wltcoeff15colorb} several SNe, namely 1990O, 1992A, 1994D,
1996X, 1998aq, 1999aw and 2000dk, that are thought to suffer negligible
extinction in their host galaxies (as compiled from the literature,
${A_V}_\textrm{host}\lesssim0.04$) are shown as crossed open circles.
Putting all of these together, 
\begin{equation}
(B-I)_\textrm{max}-E(B-I)_\textrm{MW}=(B-I)_\textrm{max}^\textrm{int}
+E(B-I)_\textrm{host},
\label{bis}
\end{equation}
one expects that their colours corrected for Galactic extinction,
$(B-I)_\textrm{max}^{\textrm{MW}_\textrm{corr}}$, are equal to the
unreddened intrinsic colours at maximum, $(B-I)_\textrm{max}^\textrm{int}$.
The best fit that provides the intrinsic $(B-I)$ colour relation for normal
SNe is given by:
\begin{equation}
(B-I)_\textrm{max}^\textrm{int}=(0.013\pm0.079)+(0.148\pm0.029)\times
\textrm{wltcoeff}_\textrm{15}. 
\label{intreq} 
\end{equation}

The extinction affects colour estimation and is a source of systematic
uncertainty. The SALT2 light-curve fitter is not meant to be able to disentangle
the potential extinction by dust in the host galaxies from intrinsic colour
variation. %This is a very important issue assumed to be partially responsible
%for some photometric correlations. 
In the case of nearby SNe the SALT2 colour law should be consistent with
observed law for extinction by dust in the Milky Way (extinction law by
\citealt{cardelli89}), but SN data point to a different colour-luminosity
relation (see e.g. \citealt{astier06}). It is interesting to note that in the
case of SNe 1999aa and 1997cn, two SNe that belong to other SN subsample
discussed above and that do not suffer strong host galaxy extinction
($A_V=0.025\pm0.020$, i.e. $0.025\pm0.021$ respectively, according to the
MLCS2k2 fits in \citealt{hicken09}), the relation given by Eqn.~\ref{intreq}
is clearly unsatisfactory. Nevertheless, this is in agreement with findings in
\citealt{wang09}, who reported two SN subsamples that favour different reddening
laws. A possible explanation for this might be the existence of two SN
populations with different intrinsic colours. 

Furthermore, expression in Eqn.~\ref{bis} and the obtained fits allow us to
estimate the host galaxy reddening, implicitly assuming that
$E(B-I)= 2.35 E(B-V)$. Adopting Eqn.~\ref{intreq} as the intrinsic colour
relation for (all) SNe, the value of $E(B-V)_\textrm{host}$  can be directly
obtained substituting expressions in Eqns.~\ref{cbi} and \ref{intreq} in
Eqn.~\ref{bis}:
\begin{equation}
 E(B-V)_\textrm{host}=(0.069\pm0.009)+(0.584\pm0.028) \times
c_\textrm{(SALT2)},
\end{equation}
including all SNe in the fit (shown as dashed line in
Fig.~\ref{ebvhostcolor}).
On the other hand, excluding SNe that do not exhibit same properties as
``normal'' SNe (represented by filled symbols in Fig.~\ref{ebvhostcolor})
from the fit and keeping only them in the second case, one gets:
\[ E(B-V)_\textrm{host}=(0.035\pm0.006)+(0.569\pm0.018) \times
c_\textrm{(SALT2)},\] i.e.
\[ E(B-V)_\textrm{host}=(0.151\pm0.007)+(0.553\pm0.023) \times
c_\textrm{(SALT2)},\]
respectively.

Again, there is a significant difference ($\sim$0.12 mag) in the reddening
estimates between these two subsamples, implying that the two SN groups
favour different reddening laws. However, this discrepancy may be due to
assumption of the single intrinsic colour relation. Accounting for different
intrinsic colour relations for the two SN subsamples may be a possible
explanation for this.
The main weakness of this analysis is the very limited number of normal SNe
that enter the fit in Eqn.~\ref{intreq}, also the lack of SN data that suffer
no (or negligible) host extinction from other subsample that could
provide an analogous intrinsic relation for intrinsically redder SNe.

\subsection{High-$z$ sample}

\begin{figure}
\centerline{
\includegraphics[width=8cm,angle=-90]{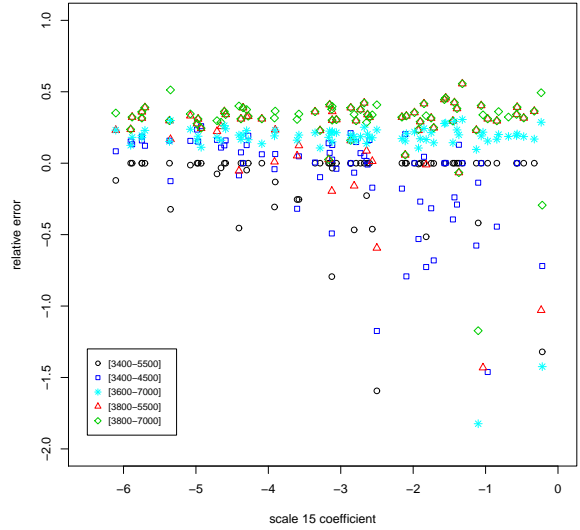}}%
\caption[]{\label{waveintervals}{\small\sl Comparison between wavelet
coefficients at scale 15 calculated on the high-$z$ SNLS3 spectra for different
wavelength intervals. The reference interval is
$3400-7000\;\textrm{\AA}$.}
\bigskip}
\end{figure}
An important question that needs to be asked is whether high-$z$ SNe follow
the same correlations as those in the previous subsection.
As an illustration, from the available set of high-$z$ spectra from SuperNova
Legacy Survey 3-year sample (SNLS3), published in \citealt{balland09},  
the wavelet analysis is applied to 94 SNe. 
%Among these, there is a
%subsample of 18 SNe, labelled as {\it ambiguous}, for which observed spectra
%are whether of poor quality, or strongly contaminated by the host galaxy
%light,
%or determined values of redshift are not reliable since the \ion{Si}{II 4000}
%feature clearly appears shifted.
Fig.~\ref{waveintervals} shows that the coefficient from the largest
scale in the DTWT, measured on the high-$z$ sample, is relatively stable to
changes in upper wavelength limit, probably because the majority of high-$z$
spectra do not reach $7000\;\textrm{\AA}$. Probing different wavelength
intervals and identifying outliers, we find a high degree of agreement when
comparing them. 
However, the wavelet coefficient at largest scale increases\footnote{and
opposite for positive values} if the lower range limit shifts
towards larger wavelengths compared to the reference one of
$3400\;\textrm{\AA}$, which might be an important issue when studying nearby
SNe.

\subsection{Stability of the largest wavelet coefficient}

\begin{figure}
\centerline{
\includegraphics[width=8cm,angle=-90]{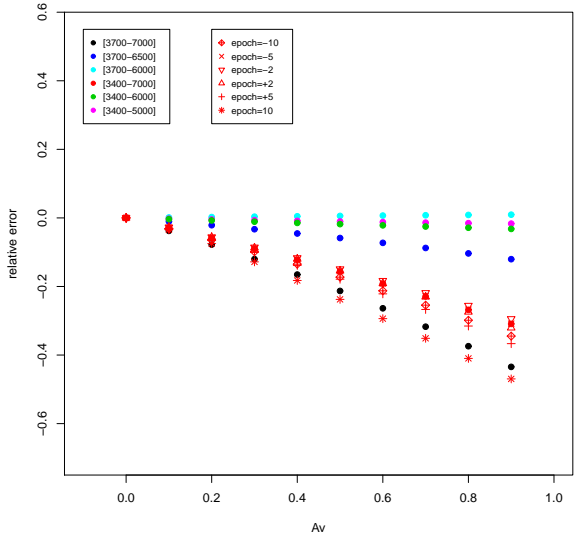}}
\caption[]{\label{waveintervalstemp}{\small\sl Comparison between wavelet
coefficients associated with the largest scale calculated on the SALT2
spectral templates for different epochs and wavelength intervals after
application of some amounts of reddening.}
\bigskip}
\end{figure}

The results in Fig.~\ref{waveintervals} are compatible with tests based on
measurements on the SALT2 spectral templates, as illustrated in 
Fig.~\ref{waveintervalstemp}. The wavelet coefficient from the coarsest scale
is tested both for stability on chosen wavelength interval and epoch relative to
$B$-band maximum. Although the dispersion seen when comparing different
measurements can be up to 50 per cent near maximum, the coefficient value
is relatively invariant across epoch over the range $[-10,+10]$ days.
Furthermore, to include the dust effects, the spectral templates in the
different wavelength intervals were reddened using the Milky Way (MW)-like
extinction curve (\citealt{cardelli89}) for different values of $R_V$ and
$E(B-V)$. It has
been shown that the effect of reddening does not strongly affect the
equivalent width of a spectral feature, especially if it is narrow (e.g.
\citealt{bronder08,arsenijevic08}). 
Similarly, despite the overall spectrum is somewhat warped, the relative error
between different cases becomes significant mainly for large extinctions
(Fig.~\ref{waveintervalstemp}). Furthermore, it can be seen that the
dependence of the largest scale wavelet coefficient on extinction effects may be
adjusted using specific wavelength intervals in the wavelet decomposition. 

\begin{figure*}
\begin{minipage}[b]{0.5\textwidth}
\centering 
\subfigure[]{ \label{coeff15colorbranchsmalla}
\includegraphics[width=8cm,angle=-90]{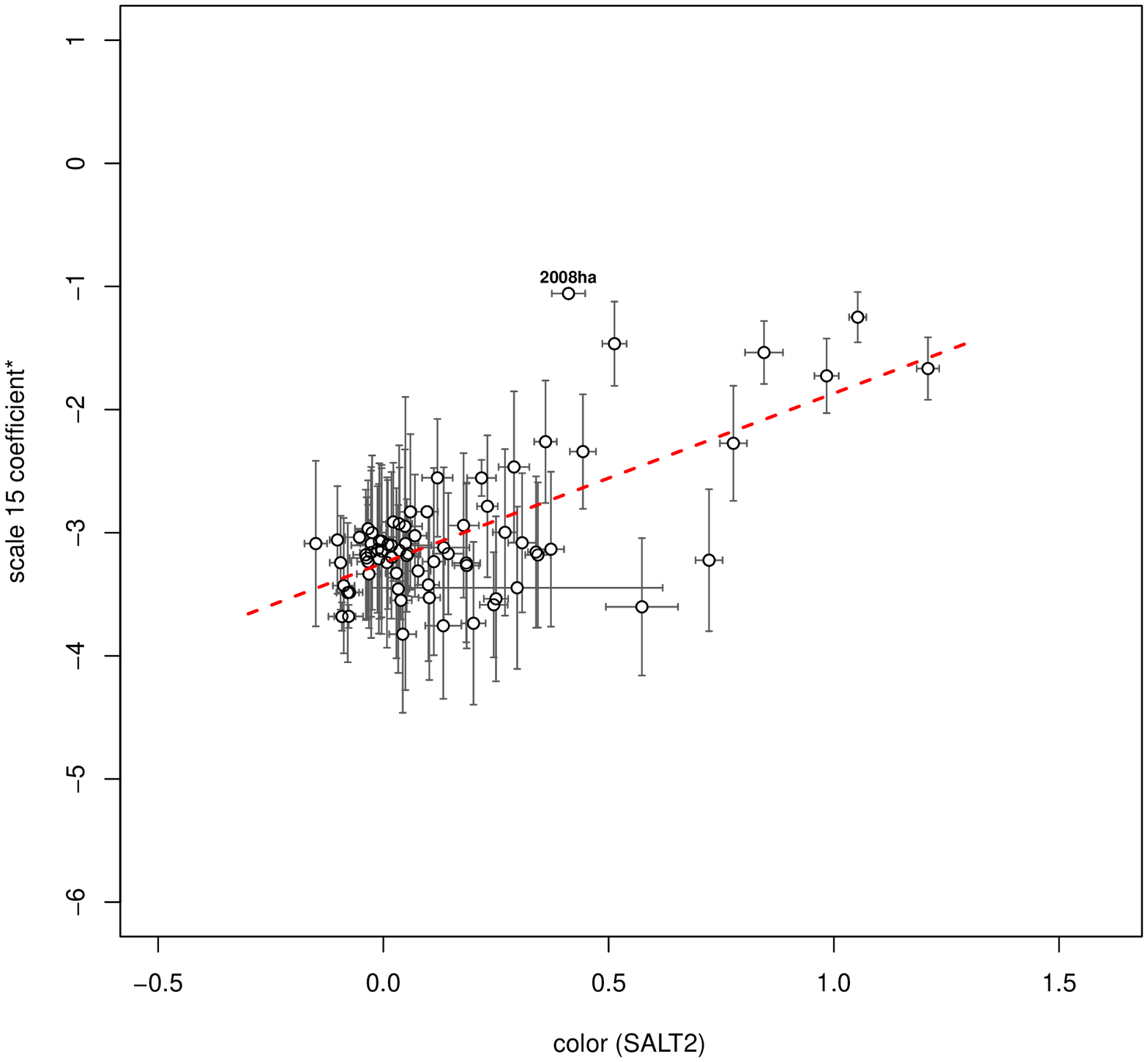}}
\end{minipage}% 
\begin{minipage}[b]{0.5\textwidth} 
\centering 
\subfigure[]{ \label{coeff15colorbranchsmallb}
\includegraphics[width=8cm,angle=-90]{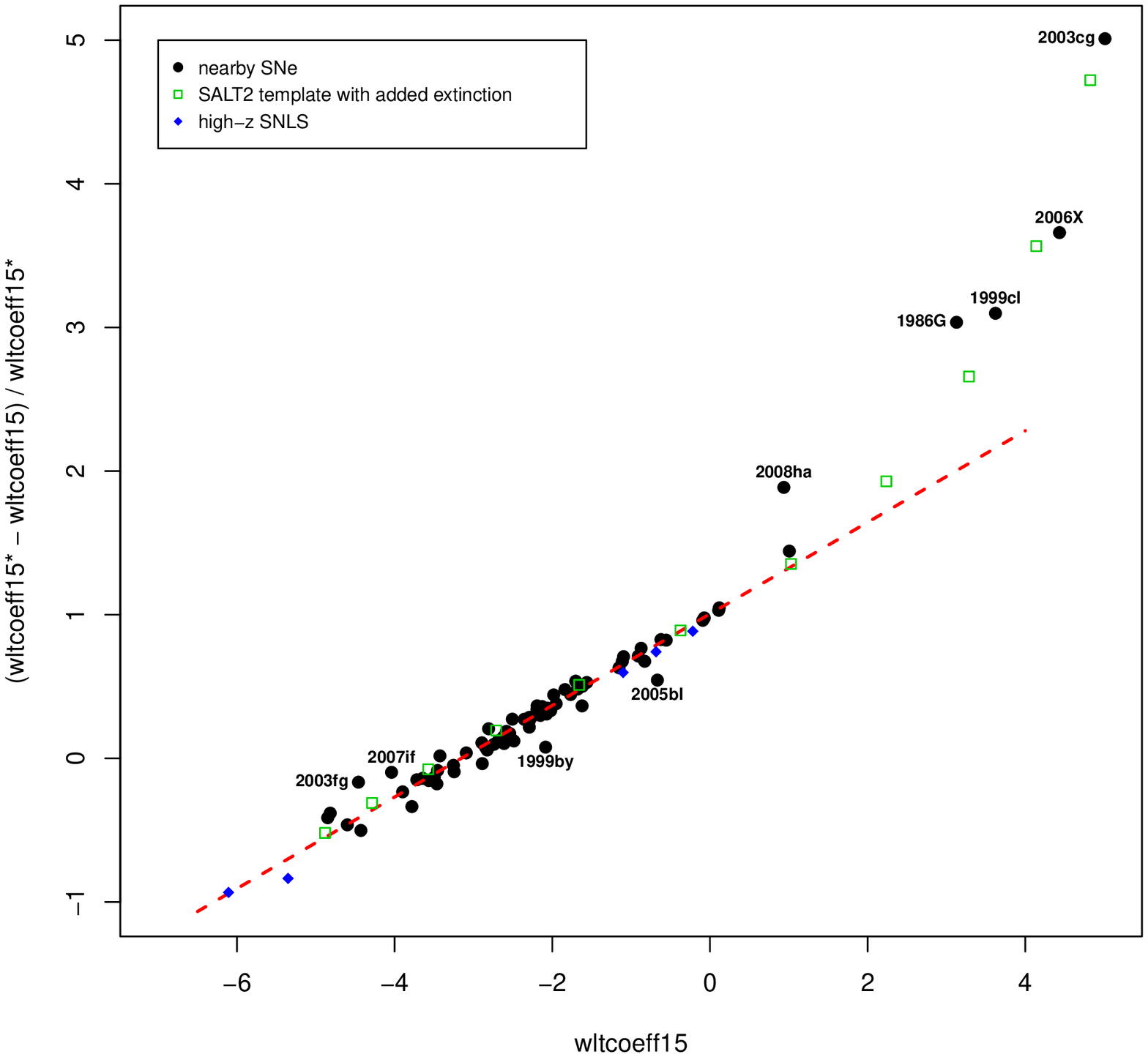}}
\end{minipage}
\caption[]{\small\sl (a) Wavelet coefficient from
the largest scale measured on the wavelet interval $[4000-6000\;\textrm{\AA}]$
versus SALT2 colour. (b) Comparison between the coefficients for
different wavelength intervals. A linear fit (dashed line) can be
used to test SNe~Ia for anomalous extinction and/or peculiarity.} 
\label{coeff15colorbranchsmall} 
\end{figure*}

To test the effects of reddening, a subset of
the ``normal'' SNe Ia was randomly chosen and some amount of reddening was
applied to them. Indeed, after obtaining new coefficients from the fits, the
differences are sufficiently large that may indicate the existence of two
populations of SNe. However, as we have seen in
Fig.~\ref{waveintervalstemp}, using the right wavelength interval, these
differences can be controlled, i.e. minimised. For instance, choosing the
intervals such as $[3700-6000\;\textrm{\AA}]$ or $[4000-6000\;\textrm{\AA}]$,
the maximum relative error reduces to a few per cent, even for large
extinction values that have been added. 
 
The previously described method was performed on the same SN spectra, using
the interval $[4000-6000\;\textrm{\AA}]$ (see
Fig.~\ref{coeff15colorbranchsmall}).
This choice guarantees the stability of the calculated values of the wavelet
coefficient to the reddening uncertainties or instrumental distortions. 

A robust regression analysis was performed to determine the best fit to the
data in Fig.~\ref{coeff15colorbranchsmalla}:
\begin{equation}
c_\textrm{(SALT2)}=
(2.36\pm0.12)+(0.73\pm0.04)\times \textrm{wltcoeff}_\textrm{15}^{*},
\label{coeff15colorbranchsmallfit}
\end{equation}
where the asterisk denotes the coefficient obtained in the interval
$[4000-6000\;\textrm{\AA}]$.

Moreover, in Fig.~\ref{coeff15colorbranchsmallb} it is demonstrated how to
perform a test that identifies SNe~Ia with anomalous extinction or strong
peculiarity. For instance, apart from four outliers in
Fig.~\ref{wltcoeff15color} (SNe 1986G, 1999cl, 2003cg and 2006X), peculiar
supernova 2008ha (extremely faint, 2002cx-like SN) strongly deviates from the
main trend despite negligible host galaxy extinction reported in
\citealt{foley09}. A few high-$z$ SNLS SNe whose spectra cover wavelengths up
to 6700 $\textrm{\AA}$ are added to the plot (given as filled diamonds) showing
an excellent agreement with nearby SNe.
 
Finally, different amounts ($0<A_V<4$, with the step $\Delta A_V=0.4$) of the
MW-like extinction are applied to the SALT2 spectral template at maximum and
obtained coefficients are plotted in Fig.~\ref{coeff15colorbranchsmallb}.
The data points follow the usual SN trend, even for the extinctions as large as
2.5 mags. Nevertheless, the figure clearly shows the two regimes, a linear
($A_V<2.5$), while there is an increase to the non-linear regime for heavily
obscured SNe with $A_V>2.5$. In addition, strongly extinguished SNe such as
1999cl, 2003cg and 2006X, are consistent with this higher power law.
After testing different extinction laws, we conclude that there is degeneracy
between different models in the linear region; the differences mainly occur at
higher values of extinction, i.e in the non-linear regime. The dependence of
this combination of wavelet coefficients shown in
Fig.~\ref{coeff15colorbranchsmallb} can be certainly used
to estimate the amount of extinction towards an individual supernova. 

\section{Discussion}

We are driven by need to grasp whether the existence of spectroscopic
dichotomy, or even a quadrotomy, recently discussed in \citealt{wang09} and
\citealt{branch09} respectively, implies either different extinction laws
or different colour evolution. As suggested in \citealt{wang05,goobar08}, the
interaction between the SN ejecta and the surrounding circumstellar shell 
with normal dust grains could explain the reported values of $R_V$ that
are lower than the standard Galactic value of $3.1$. Furthermore, the SN
spectra at early epochs are sensitive to the material in which the supernovae
are embedded.
Emitted SN light is therefore reflected by the surrounding dust clouds and
reaches the observer at different times from different directions, 
as a result of the interaction that, generally speaking, depends on the
location of the dust clouds, the density distribution of dust grains and/or the
scattering properties of dust. 
Moreover, if the dust optical depth is low, one may
assume a simple scattering scenario in which a photon escapes after scattering
off dust particle. It becomes more difficult when thick dust
shells and complex dust mixtures are considered, where multiple scatterings
usually occur. Following \citealt{patat07,blondin09}, the presence of
high-velocity features in spectra and time-variable \ion{Na}{ID} absorption
feature is good signature of influence of effects of circumstellar dust on the
SN luminosity (typical examples are SNe 1999cl and 2006X).
Besides, SN 2006X suffers significant reddening from the interstellar medium
(\citealt{wang08}). However, not all highly reddened SNe~Ia display variable
\ion{Na}{ID} features: SN 2003cg is heavily extinguished, very large
\ion{Na}{ID} equivalent width values (compared to other SNe) were measured for
this SN, but the feature remains constant over time. 
The previously described scenario can cause an effect of the light
echo around supernova, already seen, for instance, for SNe 1991T, 1995E, 1998bu,
2000cx, 2002ic, 2006X (\citealt{patat07,wang05}). Similarly, the dust shells
that surround (not necessarily highly extinguished) SNe~Ia, are reported also
for SN 2005gj\footnote{The SALT2 fit for this unusual SN is relatively poor, but
we will use the obtained estimates given in Table~\ref{nearbysnelist}.}, then
2007gi and 2007le. Indeed, all of these mentioned SNe are leaned towards the
second fit in Fig.~\ref{wltcoeff15color}. This latter supernova, recently
discovered SN 2007le (\citealt{simon09}), however, exhibits variable
\ion{Na}{ID} line but, curiously, is low extinguished, which further reflects
its intrinsic property. In addition, \citealt{yuan10} consider the
interaction between the ejecta and the circumstellar material as an alternative
manner for a supernova to generate extra luminosity, rather than any intrinsic
property of the SN. Moreover, a supernova with extremely low luminosity, SN
2008ha\footnote{Given as an open circle with high colour value, $c=0.411$ in
Fig.~\ref{wltcoeff15color}.} (\citealt{foley09}), that is spectroscopically
similar to peculiar SN
2002cx (\citealt{li03}), is in an excellent agreement with the red solid line
in Fig.~\ref{wltcoeff15color}, although it is an outlier in
Fig.~\ref{coeff15colorbranchsmall}. 
As noted in \citealt{li03}, SN 2002cx is a link between the extremes of peculiar
SNe~Ia. 
Indeed, this SN exhibits the 1991T-like premaximum spectra, but is subluminous,
such as 1991bg-like SNe. Nevertheless, the wavelet coefficient from the largest
scale reflects spectroscopic diversity of SNe~Ia, pointing to a common property
of this trend, regardless if an SN has already been classified as 1991T-,
1991bg-, 1999aa- or 2000cx-like event.

On the other hand, SNe such as 1999ee and 1989B are also found to show affinity 
for this group. These SNe exhibit a plateau in the photospheric velocity near
$B$-maximum, while normally, a standard SN Ia has very steep velocity change
from early phases towards maximum. This plateau tendency
(more precisely, a period of slowly declining velocities) coincides with the
photosphere receding through the overdense shell (\citealt{scalzo10,quimby07}).
The similar has been noticed for SNe 1991T, 1999aa, 2000cx (see e.g.
\citealt{benetti05,quimby07}), all of them having redder $(B-V)$ colour at
maximum. \citealt{quimby07} report nearly the same 
velocity evolution of the \ion{Si}{II} line for SN 2005hj, but the available
spectra for this SN do not cover wavelengths $\lambda \lesssim 3885$, therefore
the value of wavelet coefficient of interest is most likely overestimated (see
Figs.~\ref{waveintervals} and \ref{waveintervalstemp}).

Another intriguing result is that the high velocity gradient (HVG,
\citealt{benetti05}) SNe, namely SNe 1998de, 1999gd,
2002bo, 1984A, 2002bf, 1991M and 2004dt, given as light filled circles in
Fig.~\ref{wltcoeff15color} and ordered by decreasing colour, are found to 
be transitional objects, linking the two main SN trends.
A possible interpretation of this observed continuity between the
groups could be accounted as an effect of mixed extinction that these SNe
suffer, partly caused by circumstellar, but could also be due to interstellar
dust. However, another HVG SN 1981B is found in the CSM group. In addition, a
few more SNe are identified very close to these HVG SNe in direction towards
normal SNe, namely:  1998bp, 1998bu, 1999aw, 2000cf, 2001V, 2001el. Again, these
latter are characterised by whether the higher polarisation (e.g. SN 2001el), or
close resemblance to 1999aa-like or under(over)luminous SNe.
On the other hand, it has been found recently (\citealt{maeda10}) that HVG and
LVG SNe, despite the divergence in photospheric velocity, do not have
intrinsic differences. The diversity between these two supernova groups appears
as a consequence of different viewing angle from which an SN is observed. 
In other words, if one obtains a big enough SN sample, these differences will
average out and this issue will be of no concern for the cosmological use
of SNe~Ia. However, the redder SN population that satisfies relation in
Eqn.~\ref{sc15colorfitodds}, must be treated with caution because our
findings point to different intrinsic colour relations for the two SN
subsamples.

Given the previous discussion, our analysis suggests the existence of a
distinct class of SNe that is more likely governed by different physical
mechanism, i.e. considered to have a different progenitor system and/or 
explosion mechanism than standard SNe~Ia. Interestingly, SNe that support both
single- (e.g. SN 2005gj, \citealt{aldering06,prieto07}) and double-degenerate
(e.g. SN 2003fg and/or SN 2006gz, with detected envelope of unburned carbon)
scenario are found in the same group. Nevertheless, if one suspects that the
circumstellar dust is present around an SN~Ia, it is not advisable to assume a
standard interstellar extinction law, but rather should be studied it by
implementing a radiative transfer code.
The effective extinction is supposed to be generally smaller when the scattering
effects are included into consideration, since the contribution of scattered
photons makes that the total flux increases. The light that has been scattered
off the circumstellar dust cloud aims to lower the effective total to selective
extinction ratio in the optical range, completely opposite from the effect it
causes in the UV (see e.g. \citealt{wang05}). A possibility that SN colours
correlate with intrinsic brightness of SNe~Ia in such a way that this
correlation resembles reddening by dust is not rejected as well, as has been
indirectly shown in relations from Fig.~\ref{wltcoeff15color} and
Eqn.~\ref{intreq}. It is also implicitly assumed throughout that the SALT2
colour estimation is essentially perfect, despite the fact that its training
procedure is based on a sample of SNe~Ia.

Summing up, around 30 per cent of nearby SNe in our sample are found to depart 
from the main SN trend in Fig.~\ref{wltcoeff15color}. Assuming that the
similar fraction of high-$z$ population follows this tendency, a contamination
of cosmological sample by these objects could be significant. 
Recall that \citealt{wang09} recently proposed two
different $R_V$ corrections for normal and high velocity (HV) samples, which
notably improved the accuracy of distance measurements. Furthermore, the same
authors showed that the HV SNe prefer a lower value of the extinction ratio
$R_V$. On the other hand, our measurements on the spectral template show that
the largest wavelet scale coefficient is fairly stable under different
extinction laws, which suggests that this observed discrepancy has an intrinsic
character. Indeed, the two SN subsamples may have the same host galaxy
reddening law if the intrinsic colour relations were properly accounted for.

In addition, even the highly-reddened SNe are consistent with
the MW-like dust extinction behaviour in Fig.~\ref{coeff15colorbranchsmallb}.
Tests with different $R_V$ values than the canonical Galactic value of $3.1$ 
also suggest that anomalous behaviour is most likely due to high reddening.
The extinction law for the circumstellar dust from \citealt{goobar08} is also
tested and it appears very much like the curve with the MW extinction in the
linear part, while it is steeper in the non-linear regime, although for
significantly smaller values of extinction (for example, for a fixed value of
$R_V=1.8$, a CS curve that fits SN 2003cg measurements has $A_V\sim2.2$,
compared to the MW curve with $A_V\sim 2.5$). 
In order to include a more significant high-$z$ SN sample into the analysis
(compared with those few SNe that are found
in Fig.~\ref{coeff15colorbranchsmallb}), another pair of the wavelength
intervals should be studied, ones that correspond better to the rest-frame
coverage of high-$z$ SN spectra, for example between
$3000<\lambda<5000\;\textrm{\AA}$.

Wavelet scales that are directly responsible for spectral features are free of
any intrinsic color information in the input spectra, which justifies a recent
attempt to quantify SN spectral features that was done by \citealt{wagers10}.
They found correlations between various combinations of spectral features with
stretch. This is surely an important question that should be further explored.

\section{Conclusions}

The presented technique is used to test SN diversity, i.e. how significantly
the explosion of a supernova differs from widely accepted SN~Ia progenitor
scenarios, which may further lead to different intrinsic colour. For
this purpose, the wavelet transform is used to decompose SN spectra into
different scales. The largest scale coefficient is found to correlate with the
SALT2 colour parameter and long-baseline $B-I$ colour corrected for Galactic
reddening. Apart from the main trend with normal SNe~Ia, another
grouping is distinguished; its members have intrinsically redder colours when
compared to the normal ones and are recognised for various characteristics such
as the observed light echo around SN, or in addition, a signature of interaction
with circumstellar material, and/or the existence of shell-structure in or
surrounding the SN ejecta. Moreover, a few candidates of super-Chandrasekhar
mass SNe show affinity for this group. However, the questions on details on
circumstellar dust shell (or inner overdense shell), such as its mass or its
geometry, are not tackled in this work. 

This study has shown that SALT2 colour parameter can be disentangled into
intrinsic SN colour and the reddening due to dust extinction in the host
galaxy. However, with the lack of SNe from intrinsically redder subsample that
suffer negligible host galaxy extinction, the expressions for intrinsic
colour and host galaxy reddening have been found only for normal SNe.
In order to prevent the overcorrected luminosities, mentioned redder SNe should
be identified and calibrated in a different manner before being included in the
cosmology fits. 

The wavelet coefficients that were measured on certain wavelength intervals are
invariant to additional amounts of extinction that were applied to the
spectra. A combination of two largest scale coefficients from different
intervals can be further used to distinguish objects that exhibit strong
peculiarities and/or to estimate the extinction value. 

The method for additional SN sorting presented in this paper is 
applicable regardless either of the type of wavelets used in the analysis or of
the number of scales in the wavelet decomposition. Nevertheless,
there is certainly room for improvement, especially in estimating the largest
wavelet scale coefficient when the available spectra lack the full wavelength
coverage. Similarly, an additional correction could be applied to account for
spectral epochs different from the reference one at maximum light. The
implementation of the presented analysis by incorporating larger and homogeneous
spectral and photometric data will permit a deeper insight into the extinction
corrections, also the variety of the SN progenitor environments and explosion
models and will be the topic of future work.

\subsection*{Acknowledgements}
We are grateful to R. Kirshner and S. Blondin for generously allowing
us to use the CfA spectra of SN 2007af and SN 1995E before publication, also
for useful discussion. We thank P. Nugent for the valuable comments that
helped us to improve the manuscript. V. Arsenijevic acknowledges support from
FCT under grant no. SFRH/BPD/47498/2008. This work made use of the SUSPECT 
({\tt\small\urlhttp bruford\urldot nhn\urldot ou\urldot
edu/\urltilda suspect/index1\urldot html}) 
and the CfA ({\tt\small\urlhttp www\urldot
cfa\urldot harvard\urldot edu/supernova/SNarchive\urldot html}) 
SN archives.

\newpage
\onecolumn
\setlongtables
\begin{longtable}{lcrrrrrc}
\caption{List of SNe~Ia used in our study.\label{nearbysnelist}}\\
\hline\hline
SN name & Redshift & $x_1$ & $c$ & $\chi^2/\textrm{dof}$ &
$\textrm{wltcoeff}_\textrm{15}$ &$\textrm{wltcoeff}_\textrm{15}^{*}$& Refs \cr
\hline
\endfirsthead
\multicolumn{8}{l}{Table~\ref{nearbysnelist} continued}\\
\hline\hline
SN name & Redshift & $x_1$ & $c$ & $\chi^2/\textrm{dof}$ &
$\textrm{wltcoeff}_\textrm{15}$ &$\textrm{wltcoeff}_\textrm{15}^{*}$& Refs \cr
\hline
\endhead
\hline\multicolumn{8}{r}{\small\sl continued on next page}
\endfoot
\hline
\noalign {\footnotesize 
(1)~\citealt{arsenijevic08} and references therein; (2)~\citealt{matheson08};
(3)~\citealt{contreras10}; (4)~\citealt{barbon89}; (5)~\citealt{wang08};
(6)~\citealt{yamanaka09b}; (7)~\citealt{simon07};
(8)~\citealt{simon09}; (9)~\citealt{zhang10};
(10)~\citealt{perlmutter99}; (11)~\citealt{taubenberger08};
(12)~\citealt{howell06}; (13)~\citealt{pignata08};
(14)~\citealt{gomez98}; (15)~\citealt{riess98};
(16)~\citealt{thomas07}; (17)~\citealt{scalzo10};
(18)~\citealt{yuan10}; (19)~\citealt{prieto07};
(20)~\citealt{aldering06}; (21)~\citealt{foley09}.}\\
%(22)~\citealt{balland09};
\endlastfoot
1981B  &    0.0072&  $ -0.472\pm0.138$&$  0.048\pm0.038$& 1.032&$-4.428 \pm0.800$&$-2.948 \pm0.624 $&1\\ 
1984A  &   -0.0009&  $  0.051\pm0.050$&$  0.218\pm0.032$&26.152&$-1.622 \pm1.304$&$-2.555 \pm0.147 $&4\\
1986G  &    0.0018&  $ -2.226\pm0.142$&$  0.845\pm0.042$& 1.378&$  3.128\pm0.639$&$-1.536 \pm0.256  $&1\\
1989B  &    0.0024&  $ -0.913\pm0.105$&$  0.343\pm0.028$& 0.713&$-1.767 \pm1.565$&$-3.181 \pm0.589 $&1\\
1990N  &    0.0034&  $  0.996\pm0.094$&$  0.008\pm0.025$&2.025& $-1.680 \pm0.349$&$-3.242 \pm0.692 $&1\\
1990O  &    0.0307&  $ 0.450\pm0.191$& $ -0.053\pm0.032$&0.296&$ -2.740\pm0.500$&$-3.037 \pm0.058 $& 10\\
1991M  &    0.0072&  $ -1.082\pm0.239$&$  0.018\pm0.089$&0.569&$-3.254 \pm0.800$&$-3.103 \pm0.596 $&1\\
1991T  &    0.0058&  $  1.410\pm0.087$&$  0.112\pm0.025$&2.100&$  -3.718\pm0.800$&$-3.235 \pm0.760 $&1\\
1991bg &    0.0030&  $ -1.377\pm0.057$&$  0.290\pm0.034$&31.701&$ 0.118\pm0.800$&$-2.466 \pm0.614 $&1\\
1992A  &    0.0061&  $ -1.632\pm0.051$&$ -0.013\pm0.024$&1.368&$-2.115 \pm 0.326$&$-3.136 \pm0.514 $&1\\
1994D  &    0.0015&  $ -1.741\pm0.045$&$ -0.095\pm0.024$&1.258&$-2.893 \pm0.346$&$-3.244 \pm0.381 $&1\\
1994U  &    0.0048&  $ 0.000\pm7.465$& $  0.297\pm0.323$&3.341&$ -2.507\pm0.800$&$-3.447 \pm0.660 $&14, 15\\
1995E  &    0.0118&  $-0.488\pm0.122$& $  0.723\pm0.030$&0.542&$ -0.072\pm0.443$&$-3.223 \pm0.576 $&15, this work\\
1996X  &    0.0068&  $ -1.011\pm0.095$&$ -0.039\pm0.028$&0.200&$-2.582 \pm0.277$&$-3.179 \pm0.529  $&1\\
1997br &    0.0070&  $  0.369\pm0.098$&$  0.200\pm0.027$&7.670&$-0.873 \pm0.206$&$-3.736 \pm0.661 $&1\\
1997cn &    0.0162&  $ -0.550\pm0.116$&$  0.777\pm0.030$&23.903&$1.007 \pm0.800$&$-2.274 \pm0.467 $&1\\
1997do &    0.0101&  $  0.220\pm0.155$&$  0.070\pm0.026$&2.567&$ -2.020\pm0.551$&$-3.024 \pm0.497 $&2\\
1997dt &    0.0073&  $ -0.141\pm0.290$&$  0.574\pm0.080$&0.346&$ 0.111 \pm0.066$&$-3.602 \pm0.558 $&2\\
1998V  &    0.0176&  $ -0.333\pm0.087$&$ -0.004\pm0.026$&1.734&$-1.951 \pm 1.296$&$-3.149 \pm0.673 $&2\\
1998aq &    0.0037&  $ -0.353\pm0.036$&$ -0.150\pm0.025$&1.398&$ -2.848\pm0.353$&$-3.088 \pm0.672 $&1, 2\\
1998bp &    0.0104&  $ -2.440\pm0.169$&$  0.184\pm0.031$&1.108&$-1.618 \pm0.435$&$-3.246 \pm0.644 $&2 \\
1998bu &    0.0030&  $ -0.344\pm0.029$&$  0.185\pm0.027$&3.116&$-1.671 \pm0.458$&$-3.266 \pm0.674 $&1, 2\\
1998de &    0.0167&  $ -3.061\pm0.133$&$  0.443\pm0.029$&2.988&$-0.091 \pm0.684$&$-2.341 \pm0.465 $&2\\
1998dh &    0.0089&  $ -0.692\pm0.106$&$  0.055\pm0.027$&1.346&$-2.062 \pm0.123$&$-3.169  \pm0.346 $&2\\
1998dm &    0.0066&  $  0.596\pm0.251$&$  0.245\pm0.031$&0.965&$-0.621 \pm1.160$&$-3.585 \pm0.427 $&2\\
1998ec &    0.0199&  $ -0.188\pm0.323$&$  0.134\pm0.057$&0.435&$-1.153 \pm0.582$&$-3.121\pm0.652 $&2\\
1998eg &    0.0248&  $ -0.700\pm0.316$&$ -0.009\pm0.028$&0.370&$-2.151 \pm0.329$&$-3.068 \pm0.631 $&2\\
1998es &    0.0106&  $  1.011\pm0.119$&$  0.036\pm0.025$&0.553&$ -4.600 \pm1.523$&$-3.144 \pm0.674 $&2\\
1999X  &    0.0252&  $ -0.529\pm0.498$&$  0.010\pm0.050$&0.271&$-2.293 \pm0.191$&$-3.096 \pm0.524 $&2\\
1999aa &    0.0144&  $  1.170\pm0.093$&$ -0.079\pm0.025$&0.803&$-4.817 \pm1.470$&$-3.486 \pm0.566 $&1, 2\\
1999ac &    0.0095&  $  0.206\pm0.064$&$  0.029\pm0.024$&3.220&$-2.129 \pm1.388$&$-3.329 \pm0.691 $&1, 2\\
1999aw &    0.0380&  $  2.339\pm0.166$&$ -0.075\pm0.029$&1.649&$  -3.425\pm0.132$&$-3.484 \pm0.033 $&1,2\\ 
1999by &    0.0021&  $ -1.444\pm0.063$&$  0.360\pm0.025$&17.604&$-2.085 \pm1.116$&$-2.261 \pm0.497 $&1, 2\\
1999cc &    0.0313&  $ -1.582\pm0.199$&$ -0.025\pm0.031$&0.220&$-2.826 \pm0.224$&$-2.999 \pm0.627 $&2\\
1999cl &    0.0076&  $ -0.753\pm0.167$&$  0.984\pm0.027$&1.581&$ 3.622 \pm0.582$&$-1.726 \pm0.303 $&2\\
1999dq &    0.0143&  $  0.826\pm0.062$&$  0.033\pm0.025$&0.867&$-2.193 \pm2.014$&$-3.457 \pm0.682 $&2\\
1999ee &    0.0113&  $  0.722\pm0.039$&$  0.231\pm0.023$&0.690&$-2.888 \pm0.191$&$-2.785 \pm0.576 $&1\\
1999ej &    0.0137&  $ -1.400\pm0.217$&$ -0.037\pm0.038$&0.490&$-2.290\pm0.272$&$-3.206 \pm0.493 $&2\\
1999gd &    0.0185&  $ -1.082\pm0.126$&$  0.372\pm0.029$&1.829&$-0.555 \pm0.070$&$-3.134 \pm0.629 $&2\\
1999gh &    0.0077&  $ -2.329\pm0.152$&$  0.120\pm0.034$&0.381&$ -0.828 \pm0.410$&$-2.554 \pm0.478 $&2\\
1999gp &    0.0267&  $  1.749\pm0.083$&$ -0.004\pm0.018$&1.082&$ -2.539\pm 0.936$&$-3.068 \pm0.619 $&2\\
2000E  &    0.0047&  $  0.476\pm0.062$&$  0.102\pm0.024$&3.382&$-1.839 \pm0.348$&$-3.528  \pm0.668 $&1\\
2000cf &    0.0364&  $ -0.959\pm0.210$&$ -0.034\pm0.029$&0.467&$-3.245 \pm0.101$&$-2.967 \pm0.392 $&2\\
2000cn &    0.0235&  $ -2.247\pm0.209$&$  0.077\pm0.028$&0.652&$-1.562 \pm0.630$&$-3.311 \pm0.114 $&2\\
2000cx &    0.0079&  $ -1.001\pm0.042$&$ -0.088\pm0.024$&5.622&$-4.849 \pm0.207$&$-3.430 \pm0.550 $&1, 2\\
2000dk &    0.0174&  $ -1.871\pm0.144$&$ -0.032\pm0.027$&0.454&$-2.194\pm0.265$&$-3.335 \pm0.439 $&2\\
2000fa &    0.0213&  $  0.554\pm0.187$&$  0.035\pm0.029$&0.249&$-2.292 \pm0.771$&$-2.927 \pm0.637 $&2\\
2001V  &    0.0160&  $  0.988\pm0.094$&$ -0.010\pm0.020$&0.950&$ -3.091\pm0.675$&$-3.212 \pm0.607 $&2\\
2001el &    0.0039&  $ -0.245\pm0.039$&$  0.097\pm0.024$&3.477&$-2.487 \pm0.826$&$-2.830 \pm0.050 $&1\\
2002bf &    0.0242&  $ -0.298\pm0.145$&$  0.144\pm0.034$&1.087&$ -2.268 \pm0.800$&$-3.170 \pm0.493 $&1\\
2002bo &    0.0042&  $ -0.471\pm0.038$&$  0.339\pm0.018$&1.425&$-0.909 \pm0.951$&$-3.157 \pm0.615 $&1\\
2002cx &    0.0240&  $  0.351\pm0.136$&$  0.308\pm0.031$&6.960&$-2.620 \pm0.400$&$-3.082 \pm0.565 $&1\\
2002dj &    0.0094&  $  0.028\pm0.146$&$  0.060\pm0.025$&0.934&$-3.781\pm0.568$&$-2.831 \pm0.631 $&13\\
2002er &    0.0086&  $ -0.764\pm0.057$&$  0.100\pm0.024$&0.929&$-1.112 \pm1.054$&$-3.423 \pm0.620 $&1\\
2003cg &    0.0041&  $ -0.802\pm0.039$&$  1.053\pm0.019$&4.988&$5.011 \pm 0.312$&$-1.249 \pm0.204 $&1\\
2003du &    0.0064&  $  0.294\pm0.039$&$ -0.102\pm0.017$&1.342&$-2.686 \pm0.187$&$-3.058 \pm0.437 $&1\\
2003fg &    0.2437&  $  1.134\pm0.306$&$  0.043\pm0.030$&10.955&$ -4.459\pm0.500$&$-3.824 \pm0.638 $&12\\
2004S  &    0.0091&  $ -0.217\pm0.087$&$  0.022\pm0.026$&0.933&$-2.612 \pm0.800$&$-2.912 \pm0.481 $&1\\
2004dt &    0.0197&  $ -0.885\pm0.048$&$ -0.027\pm0.024$&3.012&$ -3.567 \pm1.113$&$-3.090 \pm0.594 $&1, 3\\
2004eo &    0.0157&  $ -1.044\pm0.036$&$  0.039\pm0.024$&1.098&$ -1.982\pm0.136$&$-3.549 \pm0.155 $&1\\
2005am &    0.0079&  $ -2.034\pm0.094$&$  0.133\pm0.040$&1.642&$-1.096 \pm0.800$&$-3.755 \pm0.593 $&1\\
2005bl &    0.0251&  $-2.185\pm0.100$& $  0.513\pm0.027$&7.728 &$ -0.667\pm0.653$&$-1.465 \pm0.342 $&11\\
2005cf &    0.0065&  $ -0.134\pm0.054$&$ -0.034\pm0.024$&1.119&$-2.356 \pm0.149$&$-3.235 \pm0.463 $&1\\
2005gj &    0.0593&  $ 5.000\pm0.015$& $  0.178\pm0.034$&9.610 &$ -3.464\pm0.800$&$-2.941 \pm0.587 $&19,20\\
2005hj &    0.0580&  $ 2.279\pm0.408$& $ -0.077\pm0.032$&0.778 &$ -1.703\pm0.967$&$-3.679 \pm0.094 $&1\\
2005hk &    0.0130&  $  0.044\pm0.058$&$  0.250\pm0.027$&32.085&$-2.808 \pm0.390$&$-3.537 \pm0.670 $&1\\
2006D  &    0.0097&  $ 1.341\pm0.044$& $  0.019\pm0.023$&1.288 &$ -3.645\pm0.205$&$-3.200 \pm0.164 $&16\\
2006X  &    0.0063&  $-0.258\pm0.040$& $  1.209\pm0.025$&2.411 &$  4.434\pm0.326$&$-1.667 \pm0.253 $&5,6\\
2006gz &    0.0280&  $  2.228\pm0.120$&$ -0.027\pm0.025$&1.018&$-3.898 \pm0.347$&$-3.161 \pm0.695 $&1\\
2007af &    0.0063&  $-0.431\pm0.059$& $  0.052\pm0.027$&0.545 &$ -3.455\pm0.414$&$-3.185 \pm0.457 $&7, this work\\
2007if &    0.0731&  $ 2.964\pm0.198$& $ -0.092\pm0.030$&2.180 &$ -4.039\pm0.994$&$-3.681 \pm0.598 $&17,18\\
2007le &    0.0055&  $ 0.152\pm0.054$& $  0.270\pm0.027$&1.127 &$ -2.073\pm0.247$&$-2.997 \pm0.568 $&8\\
2007gi &    0.0053&  $-1.260\pm0.117$& $  0.049\pm0.029$&1.538 &$ -3.497\pm0.621$&$-3.088 \pm0.585 $&9\\
2008ha &    0.0034&  $-3.415\pm0.353$& $  0.411\pm0.037$&1.620 &$  0.937\pm0.440$&$-1.057 \pm0.253 $&21\\

\end{longtable}

\end{document}